\documentclass[12pt,preprint]{aastex}

\newcommand\cm{{\rm cm}}
\newcommand\cs{c_{\rm s}}
\newcommand\vA{v_{\rm A}}
\newcommand\Qth{Q_{\rm th}}
\newcommand\rhocrit{\rho_{\rm crit}}

\newcommand\Msun{{\rm\,M_\odot}}

\newcommand\MJtwo{M_{\rm J,2D}}
\newcommand\kms{{\rm km\, s^{-1}}}
\newcommand\pc{{\rm\,pc}}
\newcommand\kpc{{\rm kpc}}
\newcommand\G{{\rm G}}
\newcommand\torb{t_{\rm orb} }
\newcommand\simgt{\lower.5ex\hbox{$\; \buildrel > \over \sim \;$}}
\newcommand\simlt{\lower.5ex\hbox{$\; \buildrel < \over \sim \;$}}

\slugcomment{Accepted for publication in ApJ}

\shorttitle{MRI-Driven Turbulence and Cloud Formation}
\shortauthors{Kim, Ostriker, \& Stone}

\begin{document}

\title{Magnetorotationally-Driven Galactic Turbulence and \\ 
the Formation of Giant Molecular Clouds}

\author{Woong-Tae Kim$^{1,2}$, Eve C. Ostriker$^{2,1,3}$, and 
James M. Stone$^{2,4}$}
\affil{$^1$Harvard-Smithsonian Center for Astrophysics,  \\
60 Garden St. Cambridge, MA, 02138}
\affil{$^2$Department of Astronomy, University of Maryland \\
College Park, MD 20742}
\affil{$^3$Radcliffe Institute for Advanced Study, Harvard University,
Cambridge, MA, 02138}
\affil{$^4$DAMTP, University of Cambridge, \\
Wilberforce Road, Cambridge, England CB3 9EW}

\email{wkim@cfa.harvard.edu, ostriker@astro.umd.edu, jstone@astro.umd.edu}

\begin{abstract}
Giant molecular clouds (GMCs), where most stars form,
may originate from self-gravitating instabilities in the interstellar 
medium.
Using local three-dimensional magnetohydrodynamic simulations, 
we investigate
ways in which galactic turbulence associated with the magnetorotational 
instability (MRI) may influence the formation and properties 
of these massive, self-gravitating clouds.
Our disk models are vertically stratified with 
both gaseous and stellar gravity, and subject to uniform shear corresponding
to a flat rotation curve. Initial magnetic fields are assumed to be weak and
purely vertical.  For simplicity, we adopt an isothermal equation of
state with sound speed $\cs=7\, \kms$.
We find that MRI-driven turbulence develops rapidly,
with the saturated-state Shakura \& Sunyaev parameter
$\alpha\sim(0.15-0.3)$ dominated by Maxwell stresses. 
Many of the dimensionless 
characteristics of the turbulence (e.g. the ratio of 
the Maxwell to Reynolds stresses) are similar to results from
previous MRI studies of accretion disks, hence insensitive to the 
degree of vertical disk compression, shear rate, and the presence 
of self-gravity -- although self-gravity enhances fluctuation amplitudes
slightly. The density-weighted velocity dispersions in non- or weakly
self-gravitating disks are 
$\sigma_x\sim\sigma_y\sim(0.4-0.6)\cs$ and $\sigma_z\sim(0.2-0.3)\cs$,
suggesting that MRI can contribute significantly to the 
observed level of galactic turbulence.  The saturated-state 
magnetic field strength $\bar B\sim2\mu$G is similar to typical 
galactic values.
When self-gravity is strong enough, MRI-driven high-amplitude 
density perturbations are swing-amplified to form Jeans-mass 
($\sim 10^7\Msun$) bound clouds.
Compared to previous unmagnetized or strongly-magnetized disk models, 
the threshold for nonlinear instability in the present
models occurs for surface densities at least 50\% lower,
corresponding to the Toomre parameter $\Qth\sim1.6$. 
We present evidence that self-gravitating 
clouds like GMCs formed under conditions similar to our models 
can lose much of their original spin angular momenta by magnetic braking, 
preferentially via  fields threading near-perpendicularly to their spin axes.
Finally, we discuss the present results within the larger theoretical
and observational context, outlining directions for future study.
\end{abstract}


\section{Introduction}

It has long been a challenge to understand the origins of giant
molecular clouds (GMCs).  
Observations and theory both favor ``top-down'' (instability) 
rather than ``bottom-up'' (coagulation) mechanisms,
at least for the most massive clouds ($\sim 10^5-10^6\Msun$) in which the
majority of the molecular material is found and the preponderance of
stars are born (e.g., \citealt{bli93}).  The principal difficulty with 
stochastic coagulation is that it proceeds much too slowly
to achieve GMC masses given other constraints (e.g., \citealt{bli80});
recent support for this view includes the observed lack of raw
material in small clouds within spiral arms 
\citep{hey98}, and the conclusion from
numerical simulations that magnetohydrodynamic (MHD) turbulence cannot
prevent gravitational collapse within clouds that exceed the magnetic
critical mass\footnote{$M_{\rm cr}= 1.2\times 10^4 M_\odot (B/30
  \mu\G)^3 (n_{H_2}/100\cm^{-3})^{-2}$} (e.g., \citealt{ost01,hei01}).
The latter result implies that intermediate products of coagulation
$\simgt 10^4 \Msun$, if they were created, would undergo rapid star formation
and be destroyed by the associated energy 
feedback before subsequent mergers could attain $\sim 10^6\Msun$.

The two main classes of instabilities proposed for GMC
formation are Parker and self-gravitating modes, and their hybrids
(e.g., \citealt{elm95}).  Although the azimuthal wavelengths and initial
growth rates of Parker modes suggest they could prompt cloud formation
in spiral arms (e.g., \citealt{mou74}), numerical simulations have recently
shown that nonlinear development favors short radial wavelengths and 
saturates at moderate amplitudes.  The low-density-contrast structures
that develop do not appear to be the primary progenitors of 
gravitationally-bound GMCs \citep[hereafter Paper III]{kimj00,kimj01,kim02b}.

Direct self-gravitating instabilities have similar preferred azimuthal
scales and growth rates to Parker modes, but are not self-limiting in the
nonlinear regime.
In regions {\it without} spiral structure, epicyclic motion can
temporarily conspire with shear in the swing amplifier process
\citep{too81} to yield gravitational runaway provided a sufficient
surface density. Based on two-dimensional simulations,
\citet[hereafter Paper I]{kim01} found a nonlinear instability
threshold at Toomre $\Qth=1.2-1.4$ (for definition, see eq.\ [\ref{Toomre_Q}]
below) for a range of magnetic field strengths; from three-dimensional
simulations with zero or thermal-equipartition magnetic fields, 
Paper III found $\Qth\simlt 1$.  

In high-density portions of spiral arms, local reduction or reversal of shear 
suppresses swing amplification, but magneto-Jeans instability 
(see \citealt{lyn66,elm87} and Paper I) can yield self-gravitating growth since
magnetic torques between azimuthally-adjacent regions limit the epicyclic 
motion and spin-up 
of contracting condensations. 
Numerical simulations in thin-disk models
\citep[hereafter Paper II]{kim02a} have shown that the magneto-Jeans 
instability  in spiral arms first leads to growth of gaseous spurs, 
which then fragment into Jeans-scale condensations.

The investigations of cloud formation in Papers I-III considered
either two-dimensional or three-dimensional models with zero or strong
magnetic fields (to study Parker modes). 
Under more realistic galactic
conditions with subthermal mean magnetic fields, however, new
dynamical effects are introduced by the action of the
magnetorotational instability (MRI).  Although the MRI has primarily
been studied in the context of accretion disks, 
\cite{sel99} proposed that MRI and related processes
may be important for maintaining turbulence in the outer \ion{H}{1}
disks of galaxies where the star formation rate (and supernova rate)
is low. The recent calculations of \cite{wol03} showing that a cold
phase must be present in the far outer Milky Way support the notion 
that MRI-driven turbulence prevents the settling and gravitational
fragmentation of a thin, cold gas layer that would
otherwise ensue.

To understand why the MRI is likely to be important in galaxies, it is
helpful to consider the instantaneous instability criterion for MRI modes
$\propto e^{i{\bf k \cdot x}}$, 
\begin{equation}\label{MRIcrit}
{({\bf k  \cdot}  {\bf v}_{\rm A})^2\over \Omega^2} 
{|{\bf k}|^2 \over k_z^2} < -2{d \ln\Omega  \over d\ln R} \equiv 2 q
\end{equation}
(e.g., \citealt{bal98}). Here, $\Omega(R)$ is the disk's angular rotation 
profile, and ${\bf v}_{\rm A} \equiv {\bf B}/(4\pi\rho)^{1/2}$ in 
terms of the local density $\rho$ and mean magnetic field $\bf B$.
Physically, this criterion represents the need for
(moderate) magnetic tension torques between radially-displaced fluid elements 
that are azimuthally sheared by the background flow to reinforce the
initial perturbations in angular momentum \citep{bal91}.  In a galaxy 
without strong spiral structure or in interarm regions when a 
prominent spiral pattern exists, $q \sim 1$ 
and unstable MRI modes with a variety of wavenumbers 
will be present.\footnote{E.G. using $\Omega = 26\, \kms \kpc^{-1}$ and 
scale height $H\sim 150\pc$, with 
primarily-toroidal mean $B_0\sim 1.4 \mu \G$,  and assuming wavelengths 
$\lambda_z< \lambda_R,\lambda_\theta$, eq. (\ref{MRIcrit}) requires
$\left[ {v_{{\rm A},z} H \over v_{\rm A} \lambda_z},
{ H\over \lambda_\theta} \right] \simlt 0.1$ -- which essentially calls 
for azimuthal wavelengths $\gg H$ and vertical wavelengths up to $\sim H$
(provided the vertical magnetic field is not too strong).}  

The detailed dynamical characteristics and consequences of MRI-driven
turbulence in galaxies have not previously been studied directly.
Although the literature contains many theoretical 
and numerical treatments of MRI
(e.g. \citealt{bal91,haw95,sto96,mil00,haw02}, see also 
\citealt{bal98,sto00} and references therein), these works have 
focused primarily 
on angular momentum transport in accretion disks around stars and black holes.
These accretion disks differ significantly from galactic disks in that
the latter are self-gravitating, have flat rotation curves
rather than Keplerian rotation, and are more vertically compressed by both
gaseous self-gravity and external stellar gravity rather than solely by 
a tidal stellar potential. How do the physical properties 
of a disk affect the level and characteristics of the saturated-state 
turbulence driven by MRI? 
How does this turbulence nonlinearly interact with self-gravity
to form GMC-like structures in galactic disks? 
In this paper, we address these and related
questions by performing and analyzing 
a suite of three-dimensional MHD simulations. 

As in Paper III, our disk models are local, isothermal, and vertically
stratified. Initial magnetic fields are assumed to have a weak,
uniform strength, pointing in the vertical direction (but note that
after MRI develops, $\bf B$ becomes toroidally-dominated).  The
background potential corresponds to that yielding a flat galactic
rotation curve.  In this first work on galactic MRI, we do not 
consider other features such as stellar spiral structure
and effects of the multi-phase ISM that would be present in real galaxies; 
these effects will be considered in subsequent work.

In \S 2, we briefly describe our model parameters and computational
methods. In \S3.1, we investigate the characteristics of MRI-driven
turbulence in our non- or weakly self-gravitating models, and compare
with results from previous accretion disk simulations. We directly
measure the density-weighted velocity dispersions to assess whether
MRI can be a significant source of turbulence in the diffuse ISM. We
also discuss the role played by self-gravity in supplying turbulent
energy to the ISM.  In \S3.2, we present results on $Q$ thresholds for
the formation of gravitationally bound clouds in MRI-unstable disks,
with comparison to our model results from Paper III. The properties of
the bound clouds formed and the role of magnetic fields in braking
their rotational spins is briefly discussed as well. Finally, in \S 4
we conclude and discuss our findings and related issues within a
larger context, noting outstanding questions related to galactic
turbulence and GMC formation that will be the focus of future work.

\section{Methods and Model Parameters}

For local, three-dimensional simulations of vertically stratified, 
MRI-unstable, galactic gas disks, we use the same numerical 
code and model formulation described in Paper III, with the exception that 
we modify the strength and spatial distribution of the initial magnetic field. 
We adopt an isothermal condition in both space and time, with an effective 
isothermal speed of sound $\cs=7{\,\rm\,km\,s^{-1}}$, and solve for an initial
vertical density distribution $\rho_0(z)$ in 
hydrostatic equilibrium between the thermal pressure 
gradient and the total (gaseous plus stellar) gravity. For the strength of
external stellar gravity, we define
\begin{equation}\label{s0}
s_0\equiv \left(\frac{\sigma_{*,z}\Sigma_0}{\cs\Sigma_*}\right)^2,
\end{equation}
where $\sigma_{*,z}$ and $\Sigma_*$ are the vertical velocity dispersion
and surface density of stars, respectively, while $\Sigma_0$ denotes
the gas surface density (see Paper III). To represent average
disk conditions, we take $s_0=1$ corresponding to 
$\sigma_{*,z}\approx20 {\,\rm\,km\,s^{-1}}$,
$\Sigma_*\approx 35 \,\Msun\,\pc^{-2}$, and
$\Sigma_0\approx 13 \,\Msun\,\pc^{-2}$ (e.g., \citealt{kui89,hol00}).
Note that these values give comparable stellar-disk and gas-disk vertical 
gravity at $|z|\sim H$, where $H$ stands for the gas 
disk scale height (Paper III).

We introduce an initial uniform magnetic field, 
$\mathbf{B}_0 = B_0 \mathbf{\hat{z}}$,
that points in the vertical ($\mathbf{\hat{z}}$) direction.
We parameterize its strength using 
\begin{equation}\label{beta}
\beta(z) \equiv \frac{\cs^2}{\vA^2}=\frac{4\pi\rho_0(z)\cs^2}{B_0^2},
\end{equation}
where $\vA\equiv B_0/(4\pi\rho_0)^{1/2}$ is the Alfv\'en speed.
In the numerical models presented in this paper,
we adopt $\beta(0)=100$ or 400.  The physical value of the initial 
midplane magnetic field is thus $\sim 0.2-0.4 \mu$G.
Since the density drops off with height,
high-latitude regions may have $\beta(z)<1$ with which the growth of MRI
is ineffective, but the weak midplane magnetic fields guarantee that 
the most unstable wavelength of MRI is confined within $|z|\simlt H$.
Note that although the initial magnetic field in our models is primarily
vertical (a numerical expedient that yields rapid growth of MRI), this 
condition is not inconsistent with observations of preferentially
azimuthal galactic $\bf B$-fields, because the saturated-state 
MRI has mainly toroidal $\bf B$, as we shall show in \S3.1.
Also note that 
the amplitudes of saturated-state turbulence depend on the 
net flux for pure-$z$ initial $B$-fields \citep[hereafter HGB]{haw95}; we 
will briefly discuss this in \S3.1.1 and \S4.1.
Other classes of initial magnetic field distributions that have been 
studied in non-self-gravitating MRI simulations 
(e.g., HGB; \citealt{sto96,mil00}) contain
zero net magnetic flux. Since simulations which begin with 
zero net flux take much longer to reach saturated turbulence, we
have focused on uniform vertical field models in this paper.
Further exploration of intermediate field strength and/or orientation 
may yield interesting results;  we shall, however, defer broader 
parameter surveys to future investigations.

An important measure of susceptibility to gravitational instability is the 
Toomre $Q$ parameter \citep{too64}
\begin{equation}\label{Toomre_Q}
Q \equiv \frac{\kappa\cs}{\pi G \Sigma_0}
= 1.4
        \left(\frac{\cs}{7.0 {\rm\,km\,s^{-1}}} \right)
        \left(\frac{\kappa}{36\,{\rm km\,s^{-1}\,kpc^{-1}}}\right)
        \left(\frac{\Sigma_0}{13\,\Msun\pc^{-2}}\right)^{-1},
\end{equation}
where $\kappa$ is the epicyclic frequency. 
We evolve models with $Q$ in the range of 1 to 2 to represent various
conditions in disk galaxies (e.g., \citealt{mar01}). 
Note that the often-cited critical value of $\Qth=1$ 
applies to axisymmetric 
instability in an infinitesimally thin disk. 
Even in unmagnetized cases, finite disk thickness generally reduces the 
critical $Q$ value for {\it axisymmetric}
instability to $\Qth=0.676$ for purely 
self-gravitating disk with $s_0=\infty$ \citep{gol65a,gam01} and $\Qth=0.75$ 
for a disk with $s_0=1$ (Paper III). Therefore, our model disks having
$Q\simgt1$ and $s_0=1$ are all stable to axisymmetric perturbations.

We integrate the time-dependent, compressible, ideal MHD equations in a local 
Cartesian reference frame whose center orbits the galaxy with a fixed angular 
velocity $\Omega=\Omega(R_0)$ at a galactocentric radius $R_0$. 
In this local frame, the equilibrium velocity due to galactic differential
rotation relative to the center of the box is given by 
$\mathbf{v_0}=-q\Omega x\mathbf{\hat{y}}$,
where $q\equiv-d\ln\Omega/d\ln R|_{R_0}$ is the local shear rate
and $x\equiv R-R_0$ and $y\equiv R_0 (\phi - \Omega t)$ refer to radial and 
(rotating-frame) azimuthal coordinates, respectively.
We adopt $q=1$, corresponding to a flat rotation curve.
The basic equations we solve are presented in Paper III. 
We use a modified version of the ZEUS code \citep{sto92a,sto92b} 
with a hybrid Green function/FFT solver for the Poisson equation 
(Paper III)
and applying a velocity decomposition method (Paper I) 
for less diffusive
transport of hydrodynamic variables under strong shear conditions.
For improved numerical performance in the low-density ``coronal'' region,
we employ the Alfv\'en limiter algorithm of \cite{mil00} with 
the limiting speed of the displacement current, $c_{\rm lim}=8 \cs$.
We adopt periodic boundary conditions in the $y$- and $z$-directions 
and shearing-periodic conditions at the $x$-boundaries (HGB).
From test runs with an inflow/outflow $z$-boundary, 
we have confirmed that evolution of the midplane portions of the disk is 
quite insensitive to the imposed vertical boundary conditions. 

The $(x,y,z)$ dimensions of our computational box are 
$(L_x,L_y,L_z)=(25,25,6)\times H$. Since an unmagnetized disk with $s_0=1$ 
has $H\approx 160$ pc for the solar neighborhood value 
$\Omega = 26 {\,\rm km\,s^{-1}\,kpc^{-1}}$ (Paper III), 
the total mass contained in the simulation domain is
$M_{\rm tot} = 2.1\times 10^8\Msun (\Sigma_0/13\Msun\pc^{-2})
(H/160\;{\rm pc})^2$, which is about 20 times larger than the two-dimensional
Jeans mass, 
\begin{equation}\label{mj2d}
\MJtwo\equiv \frac{\cs^4}{G^2\Sigma_0}
= 10^7 \Msun 
\left(\frac{\cs}{7\,\kms}\right)^4
\left(\frac{\Sigma_0}{13\Msun\pc^{-2}}\right)^{-1}.
\end{equation}
The numerical resolution is $128^3$ zones for all models. 
We apply initial perturbations only to the velocity using spatially 
uncorrelated, random fluctuations multiplied by the background density 
profile $\rho_0(z)$. The amplitude of the perturbations is determined
such that Max($|\delta \mathbf{v}|/\cs)=0.25$ at the disk midplane.
We will use the orbital period $\torb \equiv 2\pi/\Omega = 2.4\times 
10^8\,{\rm yrs}\, (\Omega / 26 \,{\rm km\,s^{-1}\,kpc^{-1}})^{-1}$
as the time unit in our presentation.

\section{Simulations and Results} 

To investigate MRI-driven turbulence and its interaction with self-gravity,
we have performed 7 numerical simulations. The important parameters 
and the simulation outcomes for each model are listed in Table 
\ref{tbl-para}. Column (1) gives the label for each run. Columns (2) and
(3) are the Toomre $Q$ parameter and the initial field strength in
terms of the plasma parameter at the disk midplane $\beta(0)$
(see eqs.\ [\ref{beta}] and [\ref{Toomre_Q}]). Column (4) indicates whether 
self-gravity is included in the dynamical evolution. 
Note that self-gravity in the models with ``yes'' is
slowly turned on only after turbulence becomes 
saturated, starting at $t/\torb=4.1$. 
Note also that ``no'' self-gravity models still include the initial
total gravity that maintains the unperturbed equilibrium density profile. 
Column (5) gives the final simulation outcome:
``unstable'' indicates the formation of gravitationally bound
condensations, while ``stable'' indicates fluctuating density fields
without any condensations.

Columns (6) and (7) summarize the scaling relations among the time and space 
averaged stresses and magnetic energy density in the stable models:
$R_{xy}\equiv\langle\langle\rho v_x\delta v_y\rangle\rangle$ and 
$M_{xy}\equiv\langle\langle-B_xB_y/4\pi\rangle\rangle$ are the Reynolds and
Maxwell stresses, respectively; 
$T_{xy}=R_{xy}+M_{xy}$ is the total stress;
$E_{B}\equiv\langle\langle B^2/8\pi\rangle\rangle$ is the total magnetic
energy density. Here, the double bracket $\langle\langle \, \rangle\rangle$
denotes a time- and space-average. Finally, columns (8)-(10) give
the time-averaged values of density-weighted velocity dispersions
$\langle\rho\delta v_i^2\rangle^{1/2}/\langle\rho\rangle^{1/2}$
for the stable models, where the single bracket $\langle\,\rangle$ 
denotes a space average (with the subscript $i=x$, $y$, or $z$).
For columns (6)-(10), time averages are taken typically over
2 orbits after MRI saturation. 

\subsection{Non-Collapsing Models}

In this subsection we describe the evolution of non-self-gravitating
and weakly self-gravitating models that do not form bound condensations. 
Of previous work, nonstratified simulations of HGB
with vertical initial magnetic fields, or stratified disks
of \citet[hereafter SHGB]{sto96} with sinusoidally varying fields,
are similar to our models. Both HGB and SHGB differ from the present 
models in that they used a Keplerian
rotation profile appropriate to accretion disks. We shall compare the 
results of our simulations with those of HGB and SHGB.

\subsubsection{MRI-Driven Turbulence}

We begin with non-self-gravitating model A. Figure \ref{devol} plots 
evolutionary histories of the maximum density ($\rho_{\rm max}$, solid lines) 
and density dispersion ($\langle\delta\rho^2\rangle^{1/2}$, dotted lines) of 
model A together with those of the other models. Snapshots of the perturbed 
density at four time epochs of model A are shown in Figure \ref{slice}.

The initial equilibrium of model A is a vertically stratified, shearing 
disk with $q=1$ and threaded by uniform $B_z$ with $\beta(0)=100$. 
During the initial relaxation stage ($t/\torb\simlt2$), small-scale modes 
have low-amplitude (a few percent) density fluctuations, as
shown in the perturbed density plot (Fig.\ \ref{devol}$a$). As the system 
finds the most unstable MRI mode, however, density fluctuations 
begin to grow rapidly.
The dominance of the most-unstable axisymmetric mode \citep{bal91} is evident 
in the character of the density fluctuations of Fig.\ 2{\it a,b}, which are 
correlated with the ``channel flow'' in the velocity field (\citealt{haw91}; 
see below).
Linear theory predicts that for the parameters in model A, 
the maximum growth 
rate of the MRI in model A is $\gamma_{\rm max} = \Omega/2$,
which occurs for vertical wavelength 
\begin{equation}\label{lamax}
\lambda_{\rm max}\approx 200\;{\rm pc}\;
\left(\beta(0) \over 100\right)^{-1/2} 
\left(\cs \over 7\,\kms\right) 
\left(\Omega \over 26 \,{\rm km\,s^{-1}\,kpc^{-1}} \right)^{-1}
\end{equation}
(e.g., \citealt{bal98}).
Since $\lambda_{\rm max}$ fits well within a density 
scale height of the disk 
(i.e., $\lambda_{\rm max}/2<H$), the MRI in model A 
develops similarly to the way it would in a uniform medium.\footnote{ 
The fact that $\lambda_{\rm max}/\Delta_z\sim 17$, where $\Delta_z$ is
the grid spacing in the $z$-direction, means that 
the fastest growing wavelength is numerically well resolved.}
The predicted maximum growth rate, $\gamma_{\rm max}$, marked by the 
dashed line in Figure \ref{devol}$a$, is indeed in good agreement with 
the simulation results. 

Because the fastest growing mode in model A approximates an exact nonlinear 
solution of the full MHD equations in the incompressible limit \citep{goo94}, 
it can grow to very large amplitude while preserving its essential
character. As the mode grows, the whole flow in model A 
becomes dominated by this ``channel solution,'' 
in which fluid elements in alternating horizontal layers that have 
larger (smaller) azimuthal velocity than 
the background tends to move radially outward (inward).  This inward/outward
flow drags the magnetic fields radially in directionally alternating layers, 
and the sheared azimuthal motion produces corresponding alternating layers of 
toroidal magnetic field.  As the amplitude of $v_{A,\varphi}$ becomes 
large, vertical magnetic pressure gradients compress the disk in alternating
layers.  The gas associated 
with the over-dense regions at $|z|<0.08L_z$ in Figure \ref{slice}$b$ is 
streaming radially inward, while the layers surrounding it are moving outward.
After the channel mode grows to very large amplitude, 
the nonaxisymmetric parasitic instability \citep{goo94} develops to break
up the flow into smaller-scale turbulence (Fig.\ \ref{slice}{\it c,d}).
Direct growth of nonaxisymmetric MRI  may also contribute to 
small scale kinetic and magnetic turbulence (HGB). 
At $t/\torb\simgt4$, the turbulence
has achieved a saturated state with large fluctuation amplitudes in all 
physical variables.

\begin{figure}
\epsscale{1.}
\plotone{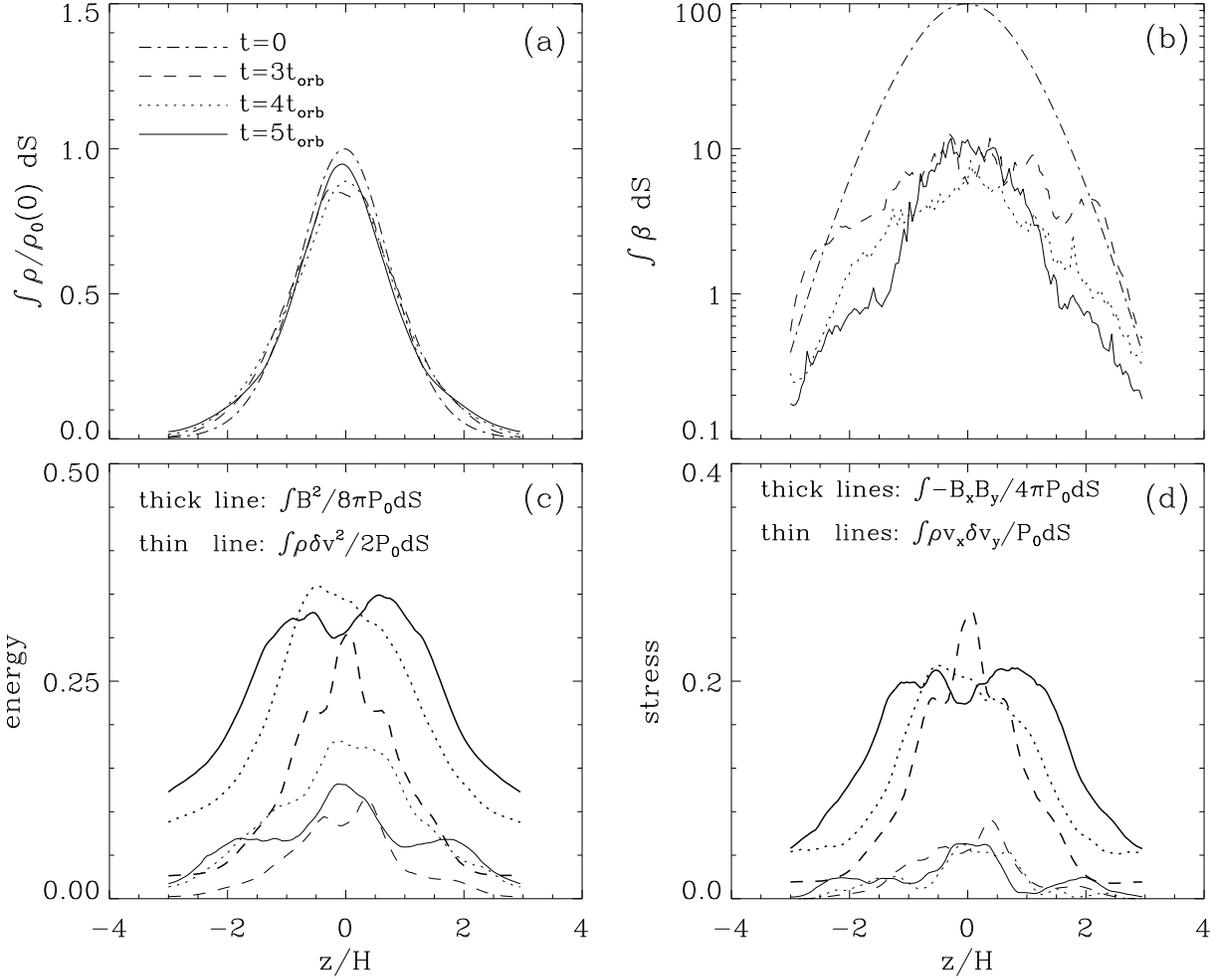}
\caption{Vertical variations of horizontally averaged quantities in model A. 
The dimensionless, differential surface element $dS$ is defined by 
$dS\equiv dxdy/(L_xL_y)$.
({\it a}) The averaged density distribution remains close to the initial 
profile
even when turbulence is fully developed. ({\it b}) In the saturated state,
$\beta$ is reduced to $\sim1-10$ at $|z|/H\simlt2$. 
In ({\it c}), the magnetic energy 
density dominates the total energy density, and in ({\it d}), 
the Maxwell stress dominates the total $x-y$ (radial-azimuthal) 
stress.
\label{zdist}}
\end{figure}

The evolution of averaged disk properties due to the MRI is illustrated in 
Figure \ref{zdist}, where we plot horizontally-averaged densities, plasma 
$\beta$, magnetic and perturbed kinetic energy densities, and Maxwell and 
Reynolds stresses of model A at times $t/t_{orb}=0,3,4,5$. 
We also list in Table \ref{tbl-A} the volume 
and time averages of various physical quantities averaged over $t/\torb=4-6$.
Magnetic fields are amplified to attain levels $\beta\sim 1-10$ at MRI 
saturation (Fig.\ \ref{zdist}$b$). The regions above $|z|=2H$ have $\beta<0.5$,
consistent with
the results of SHGB and \citet{mil00} that MRI in a stratified disk 
produces a moderately-magnetized disk midplane surrounded by a
magnetically-dominated corona. 
Overall, the magnetic energy density is dominated by the azimuthal component 
($B_y^2/8\pi$), and is larger than the perturbed kinetic 
energy density by a factor of 2 to 7, with the largest value at late times 
occurring about a scale height above the midplane 
(Fig.\ \ref{zdist}$c$).  Although the kinetic energy density is not more than
$\sim 10\%$ of the central thermal pressure and the velocity fluctuations
are subsonic at the midplane, the magnetic fields are strong enough to 
drive supersonic motions in the low-density, high-altitude regions.

Tables \ref{tbl-para} and \ref{tbl-A} 
suggest that in terms of \citet{sha73} $\alpha$ parameter describing 
the radial transport of angular momentum, the total dimensionless stress 
is $\alpha\equiv T_{xy}/P_0(0)\approx0.16$, with the Reynolds stress 
contributing only $\sim10\%$ of the total (see also Fig.\ \ref{zdist}$d$). 
This demonstrates that magnetic fields bear primary 
responsibility for mediating mass accretion in an MRI-unstable disk 
\citep{bal98}.
During the entire linear and nonlinear evolution, the averaged density 
profile in $z$ is within $20\%$ of 
the initial distribution, although the amplified magnetic 
field thickens the disk slightly via its magnetic pressure support 
(Fig.\ \ref{zdist}$a$).

\begin{figure}
\epsscale{1.}
\plotone{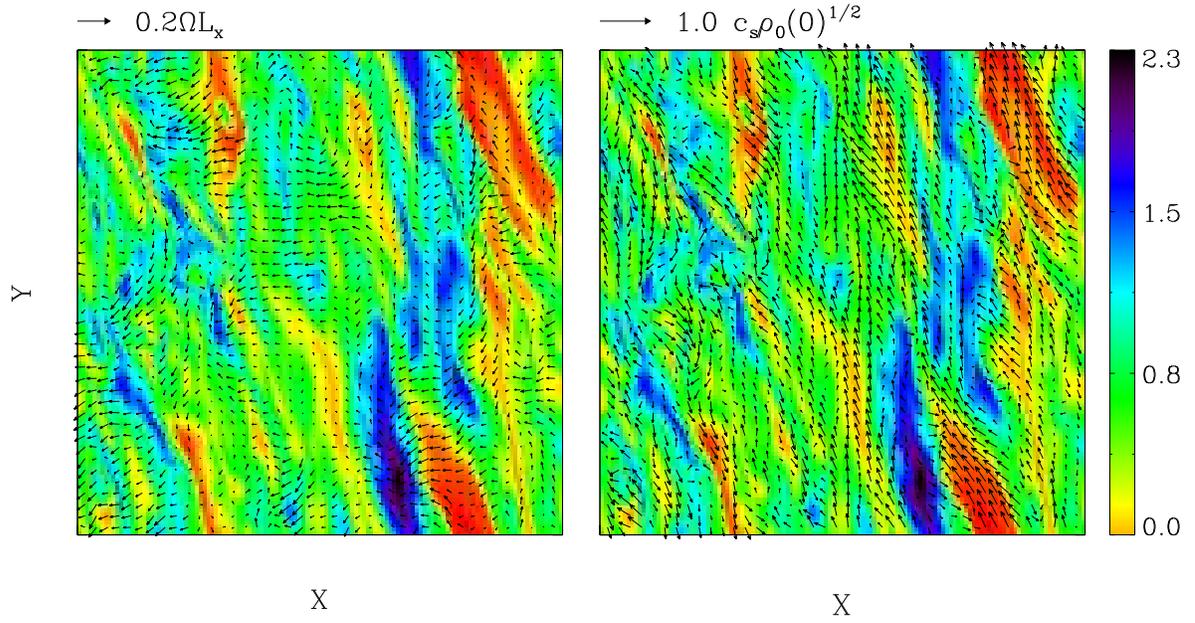}
\caption{({\it left}) Horizontal velocity and ({\it right}) 
magnetic field vectors at the midplane 
of model A at $t/\torb=5.6$ are overlaid on the maps of
midplane density ({\it greyscale}; units $\rho_0(0)$).
\label{midvec}}
\end{figure}

\begin{figure}
\epsscale{1.}
\vspace{-3cm}
\plotone{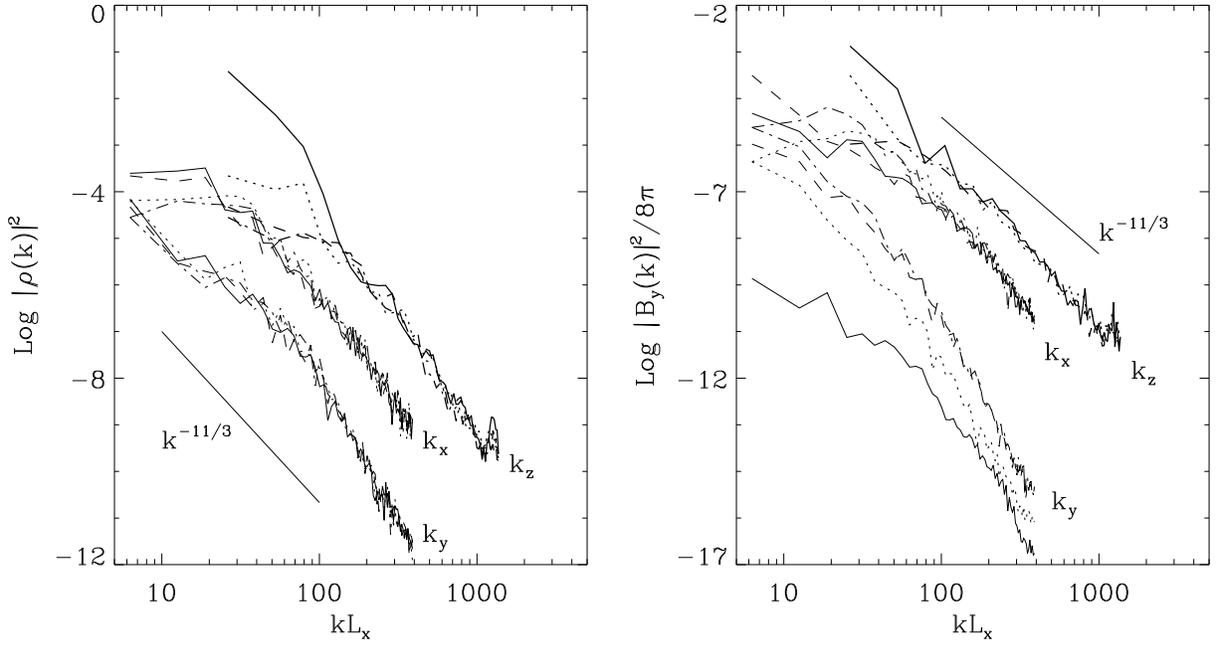}
\caption{Amplitudes of the power spectra along the $k_x$, $k_y$, and 
$k_z$ axes of density ({\it left}) 
and azimuthal magnetic field ({\it right}), averaged over $t/\torb=5-6$,
in model A. 
Along each axis, modes indicated by different lines are as follows:
for $k_x$, $n_y=n_z=0$ (solid), $n_y=1, n_z=0$ (dotted),
$n_y=0, n_z=1$ (dashed), $n_y=n_z=1$ (dot-dashed);
for $k_y$, $n_x=n_z=0$ (solid), $n_x=1, n_z=0$ (dotted),
$n_x=0, n_z=1$ (dashed), $n_x=n_z=1$ (dot-dashed);
for $k_z$, $n_x=n_y=0$ (solid), $n_x=1, n_y=0$ (dotted),
$n_x=0, n_y=1$ (dashed), $n_x=n_y=1$ (dot-dashed),
where $n_x\equiv k_x/2\pi L_x$, $n_y\equiv k_y/2\pi L_y$,
and $n_z\equiv k_z/2\pi L_z$.
For comparison, the slope of the Kolmogorov spectrum
($\sim k^{-11/3}$) is indicated in each pane.
\label{power}}
\end{figure}

In spite of the kinematic effect of the 
background shear, the density and magnetic fields are dominated by 
structure at relatively large spatial scales.
Figure \ref{midvec} displays the $x$-$y$ density structure as well as 
horizontal velocity and magnetic field vectors at the midplane of model A 
at $t/\torb=5.6$. Fourier power spectra of the density\footnote{Because the 
background density 
profile is stratified, there is a boost to the low-$k_z$ terms of the 
density power spectrum that is not associated with turbulent perturbations.}
and the azimuthal
component of the magnetic field averaged over $t/\torb=5-6$ in model A are 
plotted in Figure \ref{power}. 
Although the volume-averaged values 
of $B_x$ and $B_y$ are negligibly small (see Table \ref{tbl-A}), 
a constant-$z$ plane consists of a few magnetic domains in which $B_x$ and 
$B_y$ are distributed coherently;
across the domains, the horizontal magnetic field reverses direction.
For example,
the right panel of Figure \ref{midvec} shows that the midplane is dominated by 
negative-$B_x$ and positive-$B_y$: the midplane mean values of $B_x$ and $B_y$ 
are $-0.18$ and $0.38$ times $(4\pi P(0))^{1/2}$ at $t/\torb=5.6$, 
respectively.  Given these large mean values of $\langle B_x\rangle$ and 
$\langle B_y \rangle$ in any layer at constant $z$, the magnetic field has to
bend and reverse in the vertical direction in order to yield the small net
volume averages of Table \ref{tbl-A}  
({\it first three rows, second column}).  
This field geometry is indeed consistent with Figure \ref{power}$b$, which
shows that the maximum power of $B_y$ is attained in the 
$n_x=0=n_y$, $n_z=1$ mode; the $n_y=0$, $n_x=1=n_z$ mode is also strong.
Here, integers $n_x\equiv k_x/2\pi L_x$, $n_y\equiv k_y/2\pi L_y$, and
$n_z\equiv k_z/2\pi L_z$ denote the wavenumber in each direction.
The dominant mode for $B_y$ at late times is thus still that 
associated with channel flow.  The $B_y$ mode with next most power is 
coherent in the $y$ (azimuthal) direction and has a single reversal in the
$z$ and $x$ (radial) directions. 

Despite the fact that our model disk has lower shear and is more tightly 
vertically compressed than the models considered by HGB and SHGB, the 
characteristics of turbulence observed in the power spectra are similar. 
Among the similarities are (1) most of power is contained in modes with scales 
comparable to or larger than the disk scale height,
(2) the turbulence is highly anisotropic, as the different slopes of 
the power spectra along the different axes implies, and
(3) at well-resolved scales, the logarithmic slopes of the power spectra
are similar to the $-11/3$ Kolmogorov spectrum (see HGB; SHGB).
The kinetic-to-magnetic scaling ratios in our models are also similar 
to those in HGB. 
For instance, for a nonstratified model (Z4) with uniform magnetic fields,
HGB found $R_{xy}/M_{xy}\approx0.19$
and $T_{xy}/E_B\approx0.66$, which are very similar to
$R_{xy}/M_{xy}\approx0.13-0.14$ and $T_{xy}/E_B\approx0.64-0.65$
in models A and B of this paper (see Table \ref{tbl-para}).

The level of the turbulent fluctuations is, of course, dependent on the field 
strength and geometry, and possibly on other disk parameters.
Having uniform vertical magnetic fields, model A exhibits stronger growth
of MRI than a zero-net-$B_z$ model (IZ1) of SHGB. Table \ref{tbl-A} shows
that model A experiences fluctuations in the density with 
$\langle\delta\rho^2\rangle^{1/2}/\langle\rho\rangle\approx1.018$
and in the total pressure with $\langle\delta P_{\rm tot}^2\rangle^{1/2}
/\langle P_{\rm tot}\rangle\approx 0.629$, which are about an order of 
magnitude larger than in model IZ1 of SHGB. 
Comparing to the uniform-$B_z$ model Z4 of HGB, however, model A shows smaller
turbulent fluctuation amplitudes. 
This is because the density stratification of model A makes high-$z$ fields
effectively stronger (in terms of $\beta$), resulting in local MRI
wavelengths at high latitudes that are larger than the vertical size of box. 
As a consequence, the resulting $\alpha$ in model A (which averages over 
both high-stress and low-stress volumes) is about a half of 
$\alpha\approx0.31$ in model Z4 of HGB.

In practice, the vertical magnetic field will always be correlated in
sign on some largest radial scale, and the extent of this correlated
region -- as well as the magnitude of the vertical magnetic flux over
this scale -- will affect the amplitudes of saturated-state
turbulence.  Even over short evolutionary times 
compared to the galactic lifetime, our models show growth in the
$n_z=0$, $n_y=0$, $n_x=4$ component of the $B_z$ power spectrum,
corresponding to correlated vertical magnetic field over radial scales
comparable to the thickness of the disk.  Although realistic values
for the correlation scale and magnitude of $B_z$ are not known at
present, this evidence of ``dynamo'' activity helps to motivate our
incorporation of vertical flux in the initial conditions.  While there are 
not yet predictions from simulations of the level of vertical magnetic fields
that would develop over long timescales, one may speculate that vertical
field would build up to values similar to those we adopt, since for much
weaker vertical fields the fastest-growing MRI modes are at very small 
scales, and for much stronger vertical fields the MRI is stabilized 
(see also \S 4).

Radio observers of external face-on galaxies report line widths
of order $\sim6-10\,\kms$ in extended regions of \ion{H}{1} disks where 
no active star formation takes place \citep{dic90,van99}. 
\citet{sel99} proposed that in the absence of stellar energy inputs,
the MRI may be important in producing large-scale random motions in these 
extended \ion{H}{1} disks.\footnote{An alternative explanation for 
turbulence in extended disks without stellar energy feedback and 
in NGC 2915 attributes it to self-gravitation and bar driving \citep{wad02}; 
see also \S4.1.}
A direct way to assess this specific proposal, and more generally to 
investigate the role of the MRI in driving galactic turbulence, is to
measure the turbulent velocity dispersions from numerical simulations.
Although the current models are idealized in ways that may significantly
affect the results, it is useful to quantify the turbulence level for 
future comparisons.
As Table \ref{tbl-para} lists, the density-weighted one-dimensional
velocity dispersions in the non-self-gravitating model A are 
found to be $\sigma_x=0.45\cs$, $\sigma_y=0.43\cs$, and $\sigma_z=0.23\cs$.
In dimensional units, these correspond only to $\sim1.6-3.2\,\kms$,
much smaller than the observed line widths cited above (which however
are not purely turbulent -- see \S 4.1).
The velocity dispersions are smallest in the vertical direction, 
indicating that buoyancy does not appreciably enhance turbulence. 
Model Z4 of HGB 
reports $\sigma_x=\sigma_y=0.43\cs$ and  $\sigma_z=0.22\cs$, so that
the density stratification has little effect on the velocity dispersion.

Finally, we compare the evolution of our disk models with \citet{mil00},
who performed stratified disk models with $q=3/2$. In their simulation 
with a pure $B_z$ field, strong axisymmetric channel flows quickly emerge 
and rise buoyantly, disrupting the initial disk structure. 
The system becomes magnetically dominated everywhere at the end of the run. 
As Figure \ref{zdist} shows, however, our model A keeps its initial disk 
structure fairly intact even when the disk is highly turbulent.
High-density parts of the disk in model A have subthermal magnetic fields 
at saturation rather than becoming magnetically dominated.  In part, these
differences may be due to the differences between the circumstellar-disk
tidal gravity law $g_z= - \Omega^2 z$ and the stronger vertical gravity
provided by the gaseous and stellar disks in a galaxy.  Within a scale 
height of the midplane, the galactic gravity is a factor $\sim 2-4$ 
larger for the present models compared to $g_z= - \Omega^2 z$.  This
compresses the disk more tightly in the vertical 
direction, making the buoyant rise associated with the channel solution
less efficient. With an initial field strength having
$\beta(0)=12.5$ and a shear parameter $q=3/2$, MRI-amplified channel-flow 
magnetic fields 
in \citet{mil00} are also much stronger than those in the present models, 
contributing to the destruction of the initial disk structure.

\subsubsection{Effect of Self-Gravity on Turbulence}

We now turn to models F and G, which have self-gravity included
in their dynamical evolution but have surface densities below the
(nonlinear) gravitational instability threshold. Thus, self-gravity in
models F and G, with $Q=1.7$ and $2.0$, respectively, is not strong
enough to engender formation of bound condensations, although it certainly
enhances the fluctuation amplitudes of physical variables.  Table
\ref{tbl-F} lists the volume and time averaged quantities (over
$t/\torb=5.0-6.5$) in model F.  The fluctuations in the perturbed
density and total pressure, with
$\langle\delta\rho^2\rangle^{1/2}/\langle\rho\rangle\approx1.021$ and
$\langle\delta P_{\rm tot}^2\rangle^{1/2} /\langle P_{\rm
  tot}\rangle\approx 0.646$, are quite similar in amplitudes to those
in model A, but magnetic and perturbed kinetic energy densities in
model F are increased by about 60\%.  Note, however, that the ratios
$R_{xy}/M_{xy}$ and $T_{xy}/E_B$ are almost independent of the
presence of self-gravity (Table \ref{tbl-para}), indicating that
self-gravity, as well as the disk compression and shear parameter,
does not change the angular-momentum-transport characteristics of
MRI-driven turbulence significantly.

To gain some idea on the origin of the extra turbulent energy in the
(weakly) self-gravitating model F, we have analyzed for comparison the
data of the hydrodynamic (i.e. unmagnetized) model C (with $Q=0.7$,
$s_0=25$, and $\beta=\infty$) of Paper III.  The density-weighted
velocity dispersions, averaged over $t/\torb=1.2-2.2$, are found to be
$\sigma_x=0.12\cs$, $\sigma_y=0.08\cs$, and $\sigma_z=0.03\cs$, which
happen to be similar to $\Delta\sigma_x=0.13 \cs$,
$\Delta\sigma_y=0.07 \cs$, and $\Delta\sigma_z=0.07 \cs$, the
differences in the velocity dispersions between the MRI-unstable
models F and A. Note that the turbulence in model C of Paper III is
driven purely by hydrodynamic swing amplification (perhaps augmented by 
buoyancy effects).  Although it is not obvious how
turbulent energies of different physical origins in general ought to combine
to reach a total turbulent amplitude, 
these results suggest that motions driven by self-gravity 
{\it in a single-phase medium} can contribute only $\simlt 20\%$
of the amplitude of the velocity fluctuations that MRI can drive. Therefore,
self-gravity {\it alone}\footnote{See however the discussion in \S4.1
of gravitational driving of turbulence when thermodynamic evolution
can maintain a very low filling-factor medium.} is unlikely to be a 
significant source of turbulence in galactic disks with $Q\simgt1.7$.  
Disks with smaller $Q$ become gravitationally unstable, 
as we shall show below.

\subsection{Self-Gravitating Cloud Formation}

\subsubsection{$Q$ Threshold for Runaway and Bound Cloud Masses}

We now consider the formation of self-gravitating clouds in turbulent
galactic gas disk models.  As mentioned before, we slowly turn on
self-gravity at $t/\torb=4.1$, after MRI saturation. This slow
inclusion of self-gravity prevents an abrupt response to the
gravitational potential and enables us to explore the nonlinear
interaction of self-gravity with fully-developed turbulence.

\begin{figure}
\epsscale{1.}
\plotone{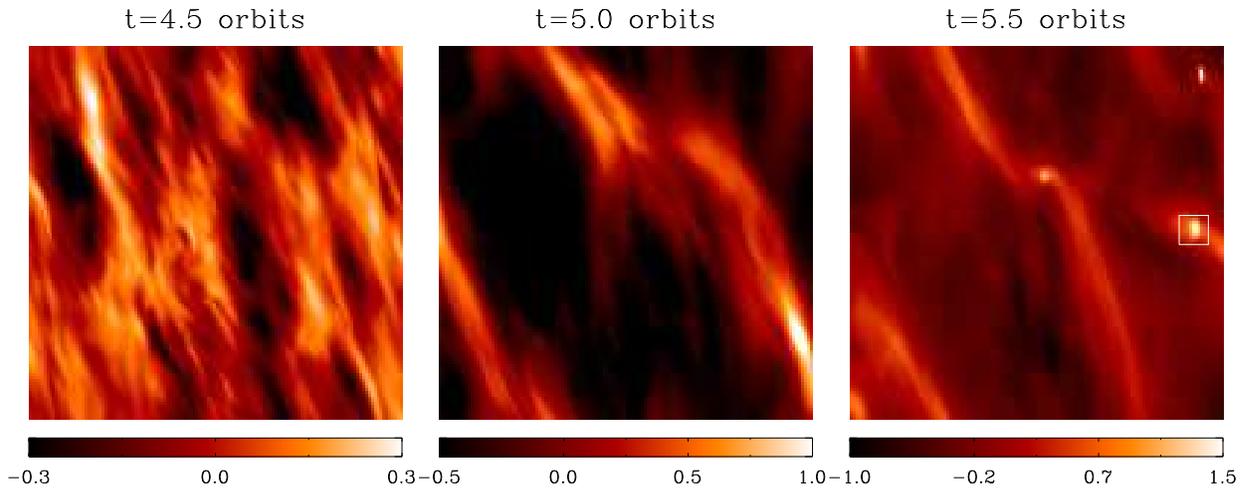}
\vspace{-2cm}
\caption{Surface density maps projected on the $x$-$y$ plane of
model C. Numbers labeling gray-scale bars correspond to
log $\Sigma/\Sigma_0$. A square in the right panel indicates the sector 
viewed as a three-dimensional visualization in Figure \ref{3Dvol}.
\label{Q10col}}
\end{figure}

As Figure \ref{devol} shows, models C-E with $Q\leq1.5$ achieve high 
enough density to produce self-gravitating runaway, while other models with 
$Q\geq1.7$ remain stable (with large fluctuations in density).
From careful examinations of density power spectra at successive times,
we identify the mechanism of self-gravitating cloud formation in models C-E 
with swing amplification of high-amplitude density perturbations. 
As discussed in the previous subsection, the power spectra of MRI-driven 
turbulence peak at small wavenumbers, so that the corresponding 
large spatial scale density perturbations can be easily swing-amplified if 
self-gravity is sufficient. 
Figure \ref{Q10col} shows surface density snapshots at $t/\torb$=4.5, 5.0, and 
5.5 of model C (with $Q=1.0$).  In the left frame,
the maximum power in the surface density is in a trailing $n_x=4, n_y=1$ mode,
but there is also considerable power in a leading mode with $n_x=-1$ and 
$n_y=1$.  The swing amplifier causes the 
leading mode to accumulate more mass as it shears into trailing 
configuration. The net result is the formation of high density filaments 
(middle frame), which collide with each other and then 
fragment to form three bound condensations (right frame). 
At their initial formation, the clumps have mass 
$M\approx\MJtwo \sim 10^7 \Msun$ each (see eq.\ [\ref{mj2d}]); 
they subsequently accrete additional
material from their surroundings. At the end of the simulation 
($t/\torb$=5.5), 
the mass contained in the three clumps is $\sim14\%$ of the total
mass, corresponding to each clump of mass  $\sim2.2\MJtwo$.

Because of its weaker self-gravity, model D with $Q=1.5$ and
$\beta(0)=100$ requires two major swing amplification events before
its eventual gravitational runaway occurs.  In particular,
intermediate-density filaments (from the first swing amplification)
have $\rho_{\rm max} \approx 4.6\rho_0(0)$ at
$t/\torb\sim 5$, but these fail to
initiate gravitational collapse. Although most of power at this point 
is in the form of trailing waves, nonlinear interactions of sheared
large-amplitude wavelets are vigorous, especially given the stresses from 
highly turbulent magnetic fields, which ends up supplying fresh 
small-$|k_x|$ modes (Paper I).  Another phase of swing amplification 
follows, this time yielding
sufficient enhancement of density to trigger gravitational runaway.
Model D forms a bound clump with $M\approx\MJtwo$ at the end of
the run.  The evolution of model E with $Q=1.5$ and $\beta(0)=400$ is
quite similar to that of model D, but the lower-amplitude saturated
state of MRI when the mean magnetic field is weaker causes the formation 
of a bound cloud to take longer.  

The self-gravitating clouds that form in our simulations turn out all to be  
magnetically supercritical. For instance, the mass-to-flux ratio of the cloud 
marked by a square in the left frame of Figure \ref{Q10col} is 
$M/\Phi_B\sim1.3\,G^{-1/2}$ at $t/\torb=5.1$ when it first appears, and
increases to $\sim3.6\,G^{-1/2}$ at $t/\torb=5.5$, 
well above the critical value of $\sim0.13\,G^{-1/2}$ (e.g., \citealt{shu92}). 
Here, $\Phi_B$ denotes the magnetic flux that passes through the cloud
and $G$ is the gravitational constant. 
The other bound clouds also have mass-to-flux ratios in the similar range,
that is, $M/\Phi_B\sim(1-4) \,G^{-1/2}$, with higher values corresponding
to more-evolved, more-massive clouds.
Note that the lower limit of this range is remarkably
similar to $M/\Phi_B\sim 0.96\,G^{-1/2}$ of a square
region projected in the $x-y$ plane 
that contains one Jeans mass in the saturated-MRI disk with
the mean $\bar B_y\sim2\mu$G. This suggests that the bound
clouds initially form by collecting material relatively 
isotropically in a way that 
preserves the magnetic flux fairly well. 
As the clouds become sufficiently self-gravitating, they begin to collapse 
and accrete more gas in all three dimensions, perhaps with accretion along 
the mean field direction enhancing 
the mass-to-flux ratio at later time. 
Although hard to quantify with our present limited resolution, magnetic 
reconnection occurring inside the clouds 
may also contribute to high mass-to-flux ratios.

As Figure \ref{devol} clearly illustrates, the threshold for nonlinear
gravitational instability in our MRI-driven turbulent galactic disk
models is in the range $Q=[1.5,1.7]$. This threshold $\Qth$ is higher
than found for either completely unmagnetized ($\Qth<1$) or
thermal-equipartition-magnetized ($\Qth\sim 1$) 3D disk models in
Paper III.  When bound clouds formed in those models, they had masses
of two to several $\MJtwo$, similar to the present results.  We note
that in our previous simulations {\it without} large-amplitude
turbulence, any model with density fluctuations as large as those in
model F (see Table \ref{tbl-F}) 
ended in gravitational runaway. Here, however, the same
turbulence that creates overdense regions can also destroy them, such that
model F is nonlinearly gravitationally stable.  The
MRI-driven turbulent disks have higher $\Qth$ than the models of Paper
III partly because the turbulence produces large-amplitude,
large-scale perturbations in the surface density (see Tables
\ref{tbl-A},\ref{tbl-F}) that reduce the effective value of $Q$ by 
$20-40\%$. In addition, 
torques from small-scale magnetic fields can drive radial angular momentum
and mass transport, much as occurs from the MRI in accretion disk
models.  Since vorticity conservation of fluid elements is the chief
barrier to gravitational instability at large enough scales in
rotating disks, radial transfer of angular momentum within/across an
overdense region can contribute to enabling runaway 
condensation (cf.\ \citealt{gam96}).

\subsubsection{Angular Momenta of Condensations}

A longstanding astronomical issue of great interest is how the total
angular momentum of a self-gravitating cloud evolves as it condenses
out of the diffuse ISM.  The prevailing idea is that magnetic fields
linking a cloud with the surrounding medium exert back-torques as it
spins up during contraction, leading to loss of angular momentum
(e.g., \citealt{gil74,mou79}). Most previous work on this ``magnetic
braking'' process has focused on the problem of an ideal cloud anchored to the
ambient medium with a prescribed magnetic field distribution, solving
to obtain the temporal evolution of the angular velocity. Using our
three-dimensional simulation data from both the present models and
those from Paper III, we will demonstrate directly in this subsection that a
cloud that condenses out of a rotating, magnetized disk can
shed significant angular momentum via magnetic braking during the
cloud-forming phase.

To describe the rotational properties of a self-gravitating cloud, we
consider the volume bounded by an isodensity surface $\rho=\rhocrit$.
We define the total angular momentum and the
total mass of a volume with $\rho>\rhocrit$ as
\begin{equation}
\mathcal{L}_z(\rhocrit)\equiv \int_{\rho>\rhocrit}
\!\!\!\!\!\rho(xv_y-yv_x) \,d^3x,
\end{equation}
\begin{equation}
\mathcal{M}(\rhocrit)\equiv \int_{\rho>\rhocrit}
\!\!\!\!\!\rho\,d^3x,
\end{equation}
respectively, where the coordinates ($x$, $y$, $z$) and the inplane velocity 
($v_x$, $v_y$) are measured relative to the center of the condensation. 
One may then calculate the mean specific angular momentum 
$J_z\equiv\mathcal{L}_z/\mathcal{M}$ as a function of $\rhocrit$.
Because model disks from which 
bound clouds originate may have different specific angular momenta
initially, it is useful to compare $J_z$ of a cloud with $J_{\rm gal}$, 
the total specific angular momentum that the cloud would have
if it did not lose any angular momentum during its formation.  For 
a region of uniform surface density, it can be shown that 
\begin{equation}\label{Jgal}
J_{\rm gal}= {\Omega \over A}\int [(1-q)x^2 + y^2] d^2x
\end{equation}
where $A$ is the area of the region in the plane of the disk
(cf.\ \citealt{mes66}).  
Thus, in the case of a flat rotation curve $q=1$, 
$J_{\rm gal}=\Omega L_0^2/12$ for a square patch of side $L_0$ in the
unperturbed disk.  For purposes of comparison, we shall take $L_0$ such 
that the total mass is the same as that within the $\rho=\rho_{crit}$ surface 
for the bound cloud that forms.

\begin{figure}
\epsscale{1.}
\plotone{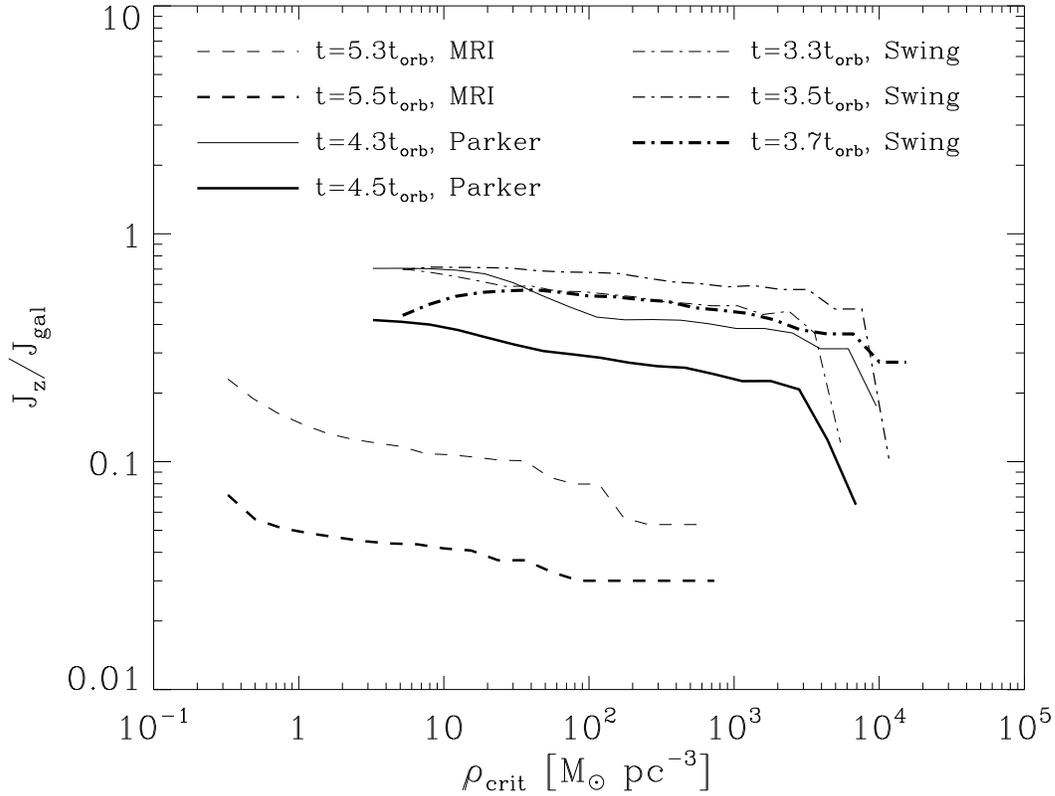}
\caption{({\it a}) Mean specific angular momenta $J_z$ inside surfaces
defined by critical density $\rho_{\rm crit}$
of self-gravitating clouds in three-dimensional simulations. 
For each cloud, $J_z$ is normalized to $J_{\rm gal}$ (see eq.\ [\ref{Jgal}]), 
the total
specific angular momentum that the cloud would retain if 
angular momentum were conserved during quasi-isotropic 
formation.
MRI indicates model C in the present work, while Parker and Swing refer 
to magnetized model D and unmagnetized model C of Paper III, respectively. 
Note that the magnetized clouds clearly lose specific angular momenta as they
collapse, while the unmagnetized model does not.
\label{angmom}}
\end{figure}

Exemplary profiles of $J_z(\rhocrit)/J_{\rm gal}$ 
for three bound clouds at a few selected times are shown in Figure
\ref{angmom}. In the figure, ``MRI'' refers to a cloud that forms
within the region marked by a square in the Figure \ref{Q10col}
snapshot of model C ($Q=1.0, s_0=1$) of this paper, while the curves
with ``Parker'' and ``Swing'' correspond to clouds formed in
magnetized model D ( $Q=0.7, s_0=25, \beta=1$) and unmagnetized model
C ($Q=0.7, s_0=25, \beta=\infty$) of Paper III, respectively.  It is
apparent that magnetized clouds (``MRI'' and ``Parker'') lower their specific
angular momentum over time as they gravitationally contract, while $J_z$ 
of an unmagnetized cloud
(``Swing'') fluctuates around $\sim0.5J_{\rm gal}$.  This implies,
consistent with expectations, 
 that
magnetic fields play a key role in removing angular momentum.
Although the resolution of the present simulations is not sufficient to 
follow this process over long periods, the characteristic angular 
momentum loss time during the interval measured is 
$J_z/|\dot J_z| \sim (1-2) \Omega^{-1}$ for the magnetized models.  
Since bound clouds typically
form  within $t_{\rm coll}\sim(1-3)\Omega^{-1}$ (see Fig.\ 
\ref{devol}), the similarity of timescales suggests that 
magnetic braking is quite efficient,
possibly enough to explain the very low observed specific angular
momenta of GMCs (e.g., \citealt{bli93}) 
compared to their corresponding $J_{\rm gal}$ values.

To help visualize the magnetic braking process that occurs in our
simulations, we show in Figure \ref{3Dvol} a volumetric rendering of an 
isodensity surface and magnetic field lines, as well as the density and 
velocity field at the midplane, for a dense clump in model C at $t/\torb=5.5$. 
The midplane velocity vectors show that the cloud is 
counter-rotating with respect to the the sense of the background shear, 
with the principal rotation axis parallel to the $z$-direction
(this sense of spin is prograde with respect to galactic rotation). 

The magnetic field lines show rather complex behavior. 
Some field lines (black) at mid-latitude regions loop 
back to the cloud, while other field lines anchor the cloud to the ambient 
medium.  Field lines drawn in green start running from the right edge of 
the box, swirl into the cloud in the counterclockwise direction, and reemerge
to swirl in the clockwise direction out to the left edge of the box.
These field lines exert net torques on the cloud surface. 
On the other hand, field lines drawn in blue that touch the ceiling of the box 
are relatively straight. The fact that the horizontal field lines
are more twisted than the vertical field lines suggests that  magnetic fields
{\it perpendicular} to the spin axis provide most of the braking. 
Although the periodic $z$-boundary conditions we have adopted tend to reduce 
vertical torsion, an important physical consideration is that there is larger 
inertia of ambient gas in the horizontal direction 
(e.g., \citealt{mou79}), tending to promote large horizontal torsion.

Finally, we remark that limited resolution in the current simulations may 
affect the measured estimate of angular momentum loss.  With higher-resolution
simulations (employing adaptive or nested meshes), it will be possible to
make a more quantitative assessment of how effective magnetic braking is
in removing a cloud's initial angular momentum.

\section{Summary and Discussion }

Understanding the origins of GMCs is key to characterizing star formation 
on a galactic scale, because the rate and mode of star formation are linked 
to the rate of cloud formation and the physical properties of GMCs at birth.  
Many different galactic-scale processes may affect GMC formation.
In this paper, we use 3D MHD simulations to initiate investigation of 
MRI-driven turbulence in disk galaxies and its role in prompting cloud-forming 
gravitational instabilities. 

Our numerical models represent local portions of shearing and vertically 
stratified gas disks with initial density profiles determined by the 
balance between thermal pressure gradients and gaseous self- and external 
stellar gravity (see \S2).
The background azimuthal velocity is set to have a local shear rate of 
$q\equiv-d\ln\Omega/d\ln R=1$, corresponding to a flat rotation curve. 
Magnetic fields are initially assumed to be purely vertical and uniform, 
with strength characterized by $\beta(0)=100$ or 400 (see eq.\ [\ref{beta}]).  
For simplicity, we have adopted an isothermal equation of state with sound 
speed $\cs=7\, \kms$; the corresponding
temperature is characteristic of the warm ISM.
We evolve disk models of varying surface density parameterized by 
the Toomre stability parameter $Q$ (see eq.\ [\ref{Toomre_Q}]); 
we consider models with $Q\sim1-2$.  We explore the nonlinear saturation of 
MRI and the growth and evolution of self-gravitating structures.
In particular, we analyze statistical properties of the turbulence,
assess the threshold for gravitational instability, and measure the masses and
angular momenta of the bound clouds that form.  Where appropriate, 
for comparison we also analyze the data from some models from Paper III.

\subsection{Properties of Turbulence}

In \S3.1 we investigate the characteristics of turbulence initiated
and sustained by MRI in non- or weakly self-gravitating models.
Despite the differences in the shear parameter, degree of vertical disk
compression, and the presence of self-gravity, the statistical
properties of the turbulence in our stratified disks are remarkably
similar to those in the nonstratified ``accretion disk''
simulations of HGB.  Our models yield saturated-state Shakura \& Sunyaev 
dimensionless $R-\varphi$ stress parameter $\alpha\sim
0.15-0.3$, ratio of the Reynolds to Maxwell stresses
$R_{xy}/M_{xy}\approx0.13$, and ratio of the total stress to
magnetic energy density $T_{xy}/E_B\approx 0.64$ (see Table
\ref{tbl-para}).  The shapes of the turbulent power spectra are 
also similar to those of HGB.  With a level
$\alpha\sim0.2$ (see also \citealt{haw02}) characterizing angular momentum
transport and $\cs=7\,\kms$, the MRI-driven gas accretion time 
$t_{\rm  acc}\sim R^2\Omega/(\alpha\cs^2)$ would exceed $10^{11}$ yr at 
the solar circle. 

To quantify the level of turbulence that is driven by MRI in our models,
we directly measure the one-dimensional velocity dispersions in each 
coordinate direction.  We find density-weighted
velocity dispersions in models with $\beta(0)=100$ are 
$\sigma_x\sim\sigma_y\sim(0.4-0.6)\cs$ and $\sigma_z\sim(0.2-0.3)\cs$,
with the mean plasma parameter $\bar\beta\sim1-2$ at saturation.
With $\cs=7\,\kms$ and $\rho_0(0)=10^{-24}$ g cm$^{-3}$,
this implies that MRI in the present models produces turbulent levels 
up to $\sigma_x\sim\sigma_y\sim4\,\kms$ and $\sigma_z\sim2\,\kms$,
and RMS magnetic field strength $\bar B\sim(1.8-2.3)\mu$G. These 
magnetic field strengths are similar to estimates in the 
Milky May and in external galaxies (e.g., \citealt{ran94,bec00,bec02}). 
The velocity dispersions are 
significantly lower than the observed line widths of $\sim7\,\kms$ and
$\sim11\,\kms$, respectively, for 
the cold and warm \ion{H}{1} components in the Solar
neighborhood (e.g., \citealt{hei03}).  Since the thermal linewidth 
from cold gas is negligible and that from the warm gas (in the stable regime) 
is $\sim7\,\kms$, the warm and cold phases both have turbulent velocity
dispersions $\sim 7\,\kms$, significantly larger than the turbulence 
levels in the present models.  For the extended disks of face-on galaxies,
radio observations of \ion{H}{1} emission linewidths imply  
$\sigma_z \sim6-10\,\kms$ (e.g., \citealt{dic90,van99,pet01}) for the 
atomic gas; in the absence of absorption measurements,  it is not directly
possible  to distinguish the turbulent and thermal contributions (or
the proportions of warm/cold gas).\footnote{Interestingly, if 
$\sigma_{\rm obs}^2= \sigma_{\rm turb}^2 + f_{\rm warm} c_{s, \rm warm}^2$ 
where $f_{\rm warm}$ is the mass fraction in the warm phase
(and assuming $\sigma_{\rm turb}$ is the same for all gas), then values of 
$\sigma_{\rm obs}$ lower than $c_{s, \rm warm} \approx 7\,\kms$ 
suggest that (1) $f_{\rm warm} <1$, i.e. at least {\it some} cold gas is 
present (consistent with theoretical expectations, cf.\ \citealt{wol03}),
and (2) $\sigma_{\rm turb}$ is lower than the ambient medium's sound speed
$c_{s, \rm warm}$.  For example, if $\sigma_{\rm turb}\approx 2\, \kms$ as 
found for $\sigma_z$ in our models, then with $c_{s, \rm warm} 
\approx 7\,\kms$, $\sigma_{\rm obs} = 6 \,\kms$ would imply 
$f_{\rm warm}=0.65$ and $f_{\rm cold}=0.35$.}

Because the level of the velocity fluctuations depends on the mean 
magnetic flux, one might argue that we could have obtained larger
$\sigma$'s, had we adopted larger initial vertical $B$-fields.
HGB found, for example, that the saturation 
magnetic energy density is linearly proportional to $\lambda_{\rm max}$ and 
thus to the mean value $B_z(0)$, or $\beta(0)^{-1/2}$.
If the MRI in a single-phase medium is to produce $\sigma_z$ up to the 
observational values, these scalings would imply that the 
vertical magnetic field should have $\beta(0)\simlt 10$ 
(assuming that the magnetic and perturbed kinetic energy densities at 
saturation are also linearly proportional). 
Equation (\ref{lamax}) would then require $\lambda_{\rm max}\simgt600$ pc 
for the vertical wavelength of MRI; if this is larger than the thickness 
of the disk, the saturation energy density would have to be reduced 
accordingly. 
Even if larger mean vertical fields were able to produce larger
velocity dispersions, the implied total magnetic energy densities would exceed
observed values in disk galaxies.
Given these constraints, we conclude that MRI acting in a 
{\it single-phase} medium could not generate the level of turbulence implied 
by \ion{H}{1} line width measurements.

It is unknown -- but of great interest -- whether MRI processes acting in 
a much less uniform, multi-phase disk could yield appreciably larger 
velocity dispersions.  One might expect, for example, a larger velocity
dispersion for a given MRI 
driving rate if larger effective mean free paths 
for dissipative 
interactions in a cloudy medium (compared to a single-phase medium) 
reduces the turbulent decay rate.
Even if the MRI angular momentum transport rate -- proportional in steady 
state to the energy dissipation rate per unit mass 
$\alpha (P/\bar \rho) q \Omega$  -- increased by a factor $4-9$ for a
factor $2-3$ increase in $\sigma_{\rm turb}$ compared to the above, 
the accretion time ($\propto 1/\alpha$) is still comparable to 
the Hubble time.  This timescale for inflow is long enough to allay the 
historical concern of overrapid accretion if turbulence taps the orbital 
energy of the ISM in a galaxy (cf.\ \citealt{spi68}). 
In addition, the implied volume heating rate would also still be 
well within the limits set by \ion{C}{2} emission from the diffuse 
ISM \citep{wol03}.
Numerical simulations will be required to decide whether MRI-driven 
turbulence in a multi-phase medium is competitive with the conventional 
turbulent energy source for the ISM, supernova shocks (e.g., 
\citealt{cox74,mck77}).
Initial work towards this goal is already underway \citep{pio03}.
Realistic theoretical assessment of the contributing turbulent 
driving processes may ultimately require simultaneous modeling, since the
dynamical effects are not necessarily independent.

Our three-dimensional simulations show that the enhancement in the 
amplitudes of velocity fluctuations by the presence of self-gravity 
is so small (less than 30\%) that self-gravity {\it alone} is unlikely to be 
a major source of galactic turbulence for a single-phase disk.
Recently, \citet{wad99} and \citet{wad02} using two-dimensional hydrodynamic 
simulations argued that self-gravity can extract sufficient 
turbulent energy from large-scale galactic rotation to maintain the observed 
level of turbulence in the ISM.  Their conclusion is based on the development 
of a turbulent velocity field with $\sigma_{\rm turb}\sim10\,\kms$ in 
thermally and gravitationally unstable flows. 
At the point a quasi-steady state is reached in the Wada et al.\ models,
most of the gas ($\simgt 80\%$) is contained in cold ($T<100$K) clouds
occupying $<1\%$ of the disk area.  Thus, the cloudy medium resembles 
a disk of collisionless particles 
which has kinetically heated itself through large-scale
gravitational instabilities until the typical dispersion in random 
velocities $\sim 10\,\kms$ is sufficient to render $Q_{\rm eff}>1$.  These
sorts of models are likely appropriate for the ISM in galactic
center regions (often dominated by molecular gas) rather than for the ISM in
the main and outer disk, where surface densities are an order of magnitude
lower than those considered by Wada et al., and where observations 
indicate that a much smaller proportion of the gas is in cold, dense clouds.

In an application to NGC 2915, \citet{wad02} also presented a 2D model
that extends up to $R=15$ kpc. They found that rapid cooling of
gas makes even very low surface-density disks highly susceptible to
thermal and gravitational instability that produces a cloudy medium with
$\sigma_{\rm turb}\sim2-4\,\kms$. Combined together with
our finding that turbulence driven by self-gravity is not significant
as long as the disks are weakly self-gravitating (with $Q\simgt1.7$)
and remain isothermal, this suggests that the dynamical importance of
self-gravity may depend strongly on the effective filling factor of
the gas.  It is therefore essential to run 3D models with a range of
properties to determine how the amplitudes of turbulence from
self-gravitational driving scale with the relative proportions of
cold, dense, and warm (or hot) diffuse gas. These proportions of gas
in different phases cannot be tuned directly since turbulence itself
can affect the ratio of dense to diffuse gas, via collisional shock
heating and other related processes.

\subsection{Bound Cloud Formation}

In \S3.2, we study directly the nonlinear interaction of MRI-driven
turbulence with self-gravity, showing that Jeans-mass-scale 
bound condensations form provided that the mean gas surface density
is sufficiently high.  In terms of the Toomre parameter, the nonlinear
instability condition is $Q<\Qth \approx 1.6$. 
The route to cloud formation is swing amplification of nonlinear density 
fluctuations ($\delta \rho/\rho \sim 2.3-4.6$) over large spatial scales 
($\sim \kpc$) 
that arise as a consequence of the MRI.\footnote{
In principle, turbulence driven by other processes -- such as
supernovae -- would also be able to produce the high-amplitude
density fluctuations needed to seed swing at $Q$-values as large as
1.5.  However, it is not yet clear whether other mechanisms can produce
density enhancements over the required $\simgt \kpc$ spatial scales.}
Compared to our previous 3D models of unmagnetized or strongly-magnetized 
($\beta=1$) disks which yielded $\Qth \simlt 1$ (Paper III), 
the MRI-unstable disks have larger $\Qth$. 
This higher $\Qth$ may partly be because self-gravity is initially strong
in the large-amplitude density fluctuations driven by turbulence,
enhancing the effectiveness of the swing amplifier; the higher $\Qth$ may
also owe in part to the ability of small-scale magnetic fields 
to transfer angular momentum between
over- and  under-dense regions locally. 

The $Q$ threshold we find is similar to the empirical result 
$\Qth \sim 1.4$ obtained by \citet{mar01} based on azimuthally-averaged 
gas surface
densities in a large sample of galaxies.  
Although intriguing, we caution that this close coincidence may 
be particular to the details of the current models (e.g. the net poloidal 
magnetic flux, the phase state of the gaseous medium).  Nevertheless, we 
consider it a significant success that theory and observation have arrived at 
essentially the same simple criterion for the onset of active star formation in
galactic disks.

Gravitationally bound clouds that form in our simulations are all
magnetically supercritical with the mass-to-flux ratio of
$M/\Phi_B\sim(1-4)G^{-1/2}$ and have typical masses of a few
$10^7\Msun$ each, corresponding roughly to the two-dimensional Jeans
mass at the mean surface density. These clouds have masses consistent
with the largest atomic or molecular clouds found in disk galaxies
(e.g., \citealt{elm83,vog88,ran90,sak99}).  The molecular components
of the largest Milky Way GMCs are nearly an order of magnitude lower
in mass.  The added contribution from \ion{H}{1}, which increases
total masses of Milky Way GMCs by at least a factor of two
\citep{bli93}, reaching $\simgt 10^7\Msun$ in some cases
\citep[e.g.,][]{elm_elm87}, accounts for some of the disparity between
our present results and observed molecular clouds.  In addition, the
disparity may owe in part to expected differences between cloud masses
in systems with and without strong spiral structure, as discussed
below.

Table \ref{tbl-sum} summarizes the model prescription/parameters and
main results for bound cloud formation from Papers I-III and the
present work.  As Table \ref{tbl-sum} indicates, our simulations so
far have considered both disk models without features in the
background gravitational potential (Paper I in 2D; Paper III and this
work in 3D), and disk models that include a background (stellar)
spiral potential (Paper II in 2D).  All of our models to date are
local.  

Although restricted to the 2D razor-thin limit, Paper II demonstrated
that when spiral structure is present, gravitational instabilities (in
particular, the MJI) form clouds preferentially in spiral arm regions,
consistent with the observations that the warmest, most massive GMCs
in the Milky Way are strongly correlated with spiral structure (e.g.,
\citealt{sol85,dam86,sol89}).  While the mean gaseous surface
densities for models that formed clouds in Paper I (without spiral
structure) and Paper II (with spiral structure) were similar, the
resulting cloud masses in Paper II were smaller by a factor $\sim 5-7$.
This difference owes to the overall compression of diffuse gas in
spiral arms, which reduces the Jeans mass $\propto \Sigma^{-1}$; the
mass collection zone is also limited by the width of the spiral arm
itself.

For a given gaseous surface density, cloud masses are also affected by
the thickness of the disk, since increasing the thickness dilutes
self-gravity, typically increasing the in-plane Jeans length by a
factor $\sim 2$.  With weaker self-gravity, a larger total area/mass
is needed to prompt instability.  Thus, the 3D models of Paper III
produced larger cloud masses at given surface density (i.e. $Q$ value)
than would have been predicted on the basis of the 2D models of Paper
I (although since instability was present only for smaller $Q$ in 3D
than in 2D, the resulting cloud masses in Papers I, III were similar).

Real galactic disks of course have finite thickness and are turbulent,
and in this sense the current models supersede the ``swing'' models of
Papers I and III.  However, real galaxies also in general contain
stellar spiral structure, which Paper II implemented only in 2D.
Finite thickness effects are expected to increase the cloud spacings
and masses that form by MJI in spiral arms compared to the results of
Paper II.  Although this could imply up to a factor $\sim 4$ increase
in cloud masses at a given surface density, turbulence could limit
this increase somewhat.  Three-dimensional simulations with explicit
modeling of spiral potential perturbations and vertical density
stratification are required to address this and related questions
directly, although our net expectations from effects currently studied
suggest total cloud masses in the neighborhood of $10^7\Msun$.

Because observed GMCs have very small specific angular momenta 
\citep{bli93} compared
to galactic values (see eq.\ [\ref{Jgal}]) on comparable mass scales, it 
is of much interest to determine when and how this
angular momentum is lost.  Using our models, we explicitly demonstrate that a
self-gravitating, contracting, magnetized cloud loses 
specific angular momentum via surface torques imposed by the 
large-scale galactic magnetic fields that thread it. 
This magnetic braking is predominantly from horizontal $B$-fields 
(perpendicular to the spin axis of the cloud) that
connect the cloud to the dense portion of the galactic disk.  
Our preliminary estimates suggest that 
the rate of angular momentum loss is sufficient to explain 
observations of low GMC spins, although higher-resolution simulations 
will be needed to confirm this result.

Finally, we remark on the mass spectrum of Galactic molecular clouds,
generally described from CO observations as a power-law 
$dN/dM\propto M^{-1.6}$ below a cutoff of a few $\times 10^6\Msun$ 
(e.g., \citealt{sol87}) (with \ion{H}{1} envelopes 
increasing individual cloud masses by factors of $2-10$). 
The gravitational instability scenario of this work and Papers I-III 
explains the upper cutoff in terms of the Jeans mass, as noted above
(see also e.g. \citealt{elm79,elm94}).  Below this cutoff,
the mass spectrum may mainly reflect the effects
of turbulent dynamics {\it within } massive (Jeans-scale) clouds as
they form and subsequently disperse.  Recent analysis of simulated models of
turbulent GMCs has shown that a power-law spectrum of moderate-density
clumps develops \citep{ost02}; the index measured is the same as the
observed Galactic cloud spectrum.\footnote{Observational clump
mass statistics within clouds are similar as well (e.g.\ \citealt{bli93}),
although apparent clumps in position-velocity space do not always
correspond to physical density concentrations, and vice versa
(e.g., \citealt{ost01}).}  If the processes that disperse a GMC
preserve the mass spectrum of former clumps as isolated clouds,
and/or if observational identification methods
decompose some large GMCs into their sub-parts, then
similar clump and cloud mass spectra are
a natural consequence of ``top-down'' GMC formation.

\acknowledgements
It is a pleasure to acknowledge valuable discussions with S.\ Balbus, 
J.\ Dickey, B.\ Elmegreen, L.\ Mundy, R.\ Narayan, F.\ Shu, and S.\ Vogel.  
We also are grateful to the referee, K.\ Wada, for a insightful 
and constructive report.
This work was supported in part by NASA grants NAG 5-9167 and 5-10780
and NSF grants AST 0205972 and AST 0307433.
Numerical simulations were performed on the O2000 system at NCSA.

\clearpage
\begin{deluxetable}{cccccccccc}
\tablecaption{Parameters of Three-Dimensional Simulations \label{tbl-para}}
\tablewidth{0pt}
\tablehead{
\colhead{\begin{tabular}{c} Model          \\ (1) \end{tabular} } &
\colhead{\begin{tabular}{c} $Q$            \\ (2) \end{tabular} } &
\colhead{\begin{tabular}{c} $\beta(0)$     \\ (3) \end{tabular} } &
\colhead{\begin{tabular}{c} \hspace{-0.5cm}Gravity                         \\ \hspace{-0.5cm}(4) \end{tabular} } & 
\colhead{\begin{tabular}{c} \hspace{-0.5cm}Results                         \\ \hspace{-0.5cm}(5)\end{tabular} } & 
\colhead{\begin{tabular}{c} \hspace{-0.5cm}$R_{xy}/M_{xy}$\tablenotemark{a}\\ \hspace{-0.5cm}(6)
\end{tabular} } & 
\colhead{\begin{tabular}{c} \hspace{-0.5cm}$T_{xy}/E_B  $\tablenotemark{b} \\ \hspace{-0.5cm}(7) \end{tabular} } & 
\colhead{\begin{tabular}{c} \hspace{-0.5cm}$\sigma_x/\cs$\tablenotemark{c} \\ \hspace{-0.5cm}(8) \end{tabular} } & 
\colhead{\begin{tabular}{c} \hspace{-0.5cm}$\sigma_y/\cs$\tablenotemark{c} \\ \hspace{-0.5cm}(9) \end{tabular} } & 
\colhead{\begin{tabular}{c} \hspace{-0.5cm}$\sigma_z/\cs$\tablenotemark{c} \\ \hspace{-0.5cm}(10)\end{tabular} }
}
\startdata
A & 1.5 & 100 &\hspace{-0.5cm} no  & \hspace{-0.5cm}...   & \hspace{-0.5cm}0.13 & \hspace{-0.5cm}0.64 & \hspace{-0.5cm}0.45 & \hspace{-0.5cm}0.43 & \hspace{-0.5cm}0.23 \\
B & 1.5 & 400 &\hspace{-0.5cm} no  & \hspace{-0.5cm}...   & \hspace{-0.5cm}0.14 & \hspace{-0.5cm}0.65 & \hspace{-0.5cm}0.21 & \hspace{-0.5cm}0.21 & \hspace{-0.5cm}0.13 \\
C & 1.0 & 100 &\hspace{-0.5cm} yes & \hspace{-0.5cm}unstable & \hspace{-0.5cm}...  & \hspace{-0.5cm}...  & \hspace{-0.5cm}...  & \hspace{-0.5cm}...  & \hspace{-0.5cm}...  \\
D & 1.5 & 100 &\hspace{-0.5cm} yes & \hspace{-0.5cm}unstable & \hspace{-0.5cm}...  & \hspace{-0.5cm}...  & \hspace{-0.5cm}...  & \hspace{-0.5cm}...  & \hspace{-0.5cm}...  \\
E & 1.5 & 400 &\hspace{-0.5cm} yes & \hspace{-0.5cm}unstable & \hspace{-0.5cm}...  & \hspace{-0.5cm}...  & \hspace{-0.5cm}...  & \hspace{-0.5cm}...  & \hspace{-0.5cm}...  \\
F & 1.7 & 100 &\hspace{-0.5cm} yes & \hspace{-0.5cm}stable   & \hspace{-0.5cm}0.14 & \hspace{-0.5cm}0.62 & \hspace{-0.5cm}0.58 & \hspace{-0.5cm}0.50 & \hspace{-0.5cm}0.30 \\
G & 2.0 & 100 &\hspace{-0.5cm} yes & \hspace{-0.5cm}stable   & \hspace{-0.5cm}0.13 & \hspace{-0.5cm}0.64 & \hspace{-0.5cm}0.58 & \hspace{-0.5cm}0.50 & \hspace{-0.5cm}0.32 \\
\enddata

\tablenotetext{a}{$R_{xy}$ and $M_{xy}$ are the $x$-$y$ components of the 
volume- and time- averaged Reynolds and Maxwell stresses, respectively.}
\tablenotetext{b}{$T_{xy}=R_{xy}+M_{xy}$, and $E_{B}$ is the volume-
and time- averaged value of the magnetic energy density.}
\tablenotetext{c}{$\sigma_x$, $\sigma_y$, $\sigma_z$ are 
the volume- and time-average of density-weighted velocity dispersions
in the $x$-, $y$-, $z$-directions, respectively.}
\end{deluxetable}

\clearpage
\begin{deluxetable}{lrcrc}
\tablecaption{Volume- and Time- Averaged Quantities in Model A
\label{tbl-A}}
\tablewidth{0pt}
\tablehead{
\colhead{Quantity}                                  &
\colhead{$\langle\langle f\rangle\rangle$}          &
\colhead{$\langle\langle \delta f^2\rangle\rangle^{1/2}$}          &
\colhead{min $f$}          &
\colhead{max $f$}
}
\startdata
$B_x/(4\pi P_0(0))^{1/2}$  &2.39$\times10^{-7}$  & 0.326 &$-1.362$ & 1.317 \\
$B_y/(4\pi P_0(0))^{1/2}$  &$-1.30\times10^{-4}$ & 0.576 &$-1.550$ & 1.467 \\
$B_z/(4\pi P_0(0))^{1/2}$  &   0.1000 & 0.174 & $-0.898$ & 1.189 \\
$B_x^2/8\pi P_0(0)$        &   0.0532 & 0.072 &   0.000 & 0.957 \\
$B_y^2/8\pi P_0(0)$        &   0.1808 & 0.147 &   0.000 & 1.218 \\
$B_z^2/8\pi P_0(0)$        &   0.0196 & 0.035 &   0.000 & 0.691 \\
$-B_xB_y/4\pi P_0(0)$      &   0.1435 & 0.168 &$-$0.880 & 1.323 \\
$-B_xB_z/4\pi P_0(0)$      &$-$0.0012 & 0.075 &$-$0.850 & 0.838 \\
$-B_yB_z/4\pi P_0(0)$      &   0.0031 & 0.110 &$-$0.885 & 0.913 \\
$\rho v_x^2/2P_0(0)$       &   0.0334 & 0.062 &   0.000 & 1.557 \\
$\rho \delta v_y^2/2P_0(0)$&   0.0293 & 0.065 &   0.000 & 1.687 \\
$\rho v_z^2/2P_0(0)$       &   0.0116 & 0.023 &   0.000 & 0.718 \\
$\rho v_x\delta v_y/P_0(0)$&   0.0180 & 0.088 &$-$1.011 & 1.760 \\
$\rho v_xv_z/P_0(0)$       &   0.0005 & 0.050 &$-$1.250 & 0.883 \\
$\rho\delta v_yv_z/P_0(0)$ &   0.0008 & 0.050 &$-$1.050 & 1.090 \\
$P_{\rm tot}/P_0(0)$       &   0.5863 & 0.396 &   0.029 & 2.478 \\
$\rho/\rho_0(0)$           &   0.3320 & 0.338 &   0.000 & 2.184 \\
$(\rho-\rho_0)/\rho_0(0)$  & 0.0000 & 0.184 &$-$0.851 & 1.233 \\
$\Sigma/\Sigma_0$          &   1.0000 & 0.201 &   0.474 & 1.779 \\
\enddata
\end{deluxetable}

\clearpage
\begin{deluxetable}{lrcrc}
\tablecaption{Volume- and Time- Averaged Quantities in Model F
\label{tbl-F}}
\tablewidth{0pt}
\tablehead{
\colhead{Quantity}                                  &
\colhead{$\langle\langle f\rangle\rangle$}          &
\colhead{$\langle\langle \delta f^2\rangle\rangle^{1/2}$}          &
\colhead{min $f$}          &
\colhead{max $f$}
}
\startdata
$B_x^2/8\pi P_0(0)$        &   0.0767 & 0.108 &   0.000 & 1.879 \\
$B_y^2/8\pi P_0(0)$        &   0.3138 & 0.215 &   0.000 & 2.076 \\
$B_z^2/8\pi P_0(0)$        &   0.0230 & 0.045 &   0.000 & 0.966 \\
$-B_xB_y/4\pi P_0(0)$      &   0.2252 & 0.249 &$-$1.378 & 2.538 \\
$-B_xB_z/4\pi P_0(0)$      &$-$0.0008 & 0.104 &$-$1.351 & 1.412 \\
$-B_yB_z/4\pi P_0(0)$      &   0.0030 & 0.164 &$-$1.468 & 1.483 \\
$\rho v_x^2/2P_0(0)$       &   0.0566 & 0.117 &   0.000 & 3.088 \\
$\rho \delta v_y^2/2P_0(0)$&   0.0412 & 0.111 &   0.000 & 3.398 \\
$\rho v_z^2/2P_0(0)$       &   0.0167 & 0.033 &   0.000 & 0.888 \\
$\rho v_x\delta v_y/P_0(0)$&   0.0310 & 0.156 &$-$2.307 & 3.544 \\
$\rho v_xv_z/P_0(0)$       &   0.0023 & 0.082 &$-$1.491 & 1.928 \\
$\rho\delta v_yv_z/P_0(0)$ &   0.0013 & 0.074 &$-$1.669 & 1.790 \\
$P_{\rm tot}/P_0(0)$       &   0.7461 & 0.482 &   0.046 & 5.474 \\
$\rho/\rho_0(0)$           &   0.3320 & 0.399 &   0.000 & 5.068 \\
$(\rho-\rho_0)/\rho_0(0)$  &   0.0000 & 0.292 &$-$0.962 & 4.093 \\
$\Sigma/\Sigma_0$          &   1.0000 & 0.410 &   0.310 & 3.493 \\
\enddata
\end{deluxetable}

\clearpage
\begin{deluxetable}{cccccc}
\tabletypesize{\small}
\tablecaption{Summary of Papers I-III and the present work \label{tbl-sum}}
\tablewidth{0pt}
\tablehead{
\colhead{paper} &
\colhead{\begin{tabular}{c} disk       \\ geometry             \end{tabular}} &
\colhead{\hspace{-0.5cm} \begin{tabular}{c} background \\ feature              \end{tabular} \hspace{-0.5cm} } &
\colhead{\begin{tabular}{c} magnetic   \\ field                \end{tabular}} &
\colhead{\begin{tabular}{c} instability  threshold \\ and outcome
                       \tablenotemark{a} \end{tabular}} &
\colhead{\begin{tabular}{c} physical   \\ mechanism            \end{tabular}}
}
\startdata
paper I & razor thin & none  & azimuthal & 
                   $Q_c=1.2-1.4$ & 
                   high shear: Swing \\
        & disks (2D) &       & $(\beta_y=\infty, 10, 1)$ & 
                   $M_{\rm cloud}$ = a few $10^7\Msun$ &
                   low shear: MJI   \\
       \\ 
paper II & razor thin & spiral & azimuthal & 
                   $M_{\rm cloud} = 4 \times 10^6\Msun$ &
                   MJI inside  \\
         & disks (2D) & arms  & $(\beta_y=\infty, 10, 1)$ & 
                                & 
                   spiral arms  \\
       \\ 
paper III & finite thickness & none  & azimuthal &
                   $Q_c \simlt 1.0$              & 
                   Swing; MJI with  \\
          & disks (3D) &       & $(\beta_y=\infty, 1)$ &
                   $M_{\rm cloud}$ = a few $10^7\Msun$ & 
                   Parker-driven boost \\
       \\ 
this paper & finite thickness & none  & vertical &
                   $Q_c =  1.6$              &  
                   Swing of MRI-driven  \\
           & disks (3D) &       & $(\beta_z=100, 400)$ &
                   $M_{\rm cloud}$ = a few $10^7\Msun$ & 
                   turbulence    \\
\enddata

\end{deluxetable}

\clearpage
\begin{figure}
\epsscale{1.}
\caption{Time evolution of maximum density (solid lines) and 
density dispersion (dotted lines) of ({\it a}) $\beta=100$ 
and ({\it b}) $\beta=400$ models. The dashed line in ({\it a})
represents the linear growth rate of the most unstable MRI mode for the
parameters of non-self-gravitating model A.
\label{devol}}
\end{figure}

\clearpage
\begin{figure}
\epsscale{1.}
\caption{Gray-scale images of the perturbed density $\delta\rho/\rho_0(0)$
at four time epochs from model A. 
Each frame (a)-(d) displays slices at 
$y=-L_y/2$ ({\it left; $x-z$ plane}) and $x=-L_x/2$ ({\it right; 
$y-z$ plane}).  The MRI ``channel solution'' that
appears strongly at $t/\torb=3$ ({\it frame [b]}) breaks up 
and generates MHD turbulence 
at $t/\torb\simgt4$ ({\it frames [c],[d]}). 
\label{slice}}
\end{figure}

\clearpage

\clearpage

\clearpage

\clearpage

\clearpage

\clearpage
\begin{figure}
\epsscale{1.}
\vspace{-2cm}
\caption{Perspective visualization of an isodensity surface 
($\rho/\rho_0(0)=10$) and selected magnetic field lines in model C
at $t/\torb=5.5$. The region fits in the projected square marked in 
Figure \ref{Q10col}, with a vertical extent of $|z|<0.04L_z$. The magnetic
field lines in green run from the right edge of the box into
the clump and from the clump to the left edge, while blue indicates field 
lines that extend to the ceiling of the box. Field lines that return
back to the clump are drawn in black. The midplane density slice is 
shown in colorscale at the bottom of the box (colorbar labels 
$\log\rho/\rho_0(0)$); 
horizontal velocity vectors in the same plane are also drawn.
\label{3Dvol}}
\end{figure}

\end{document}